\documentclass{tPHM2e}

\usepackage{graphicx}
\usepackage{amssymb}

\newcommand{\eq}[1]{Eq.~(\ref{#1})}
\newcommand{\fig}[1]{Fig.~\ref{#1}}
\newcommand{\sect}[1]{Sect.~\ref{#1}}
\providecommand{\onlinecite}[1]{\cite{#1}}

\renewcommand{\vec}[1]{\boldsymbol{#1}}

\begin{document}

\doi{10.1080/14786430903292373}
\issn{1478-6443}
\issnp{1478-6435}
\jvol{89} \jnum{34-36} \jyear{2009} \jmonth{1-21 December}
\articletype{}
\setcounter{page}{3371}
\endpage{3391}
\received{Received 15 July 2009; final version received 27 August 2009}

\title{
  Composition-dependent interatomic potentials: \\
  A systematic approach to modelling multicomponent alloys
}

\author{
  B. Sadigh$^{\rm a}$,
  P. Erhart$^{\rm a}$,
  A. Stukowski$^{\rm b}$,
  and A. Caro$^{\rm a}$
  \\
  $^{\rm a}$ {\em Condensed Matter and Materials Division, Lawrence Livermore\\ National Laboratory, Livermore, CA};
  $^{\rm b}$ {\em Institut f\"ur Materialwissenschaft, \\Technische Universit\"at Darmstadt, Germany}
}
\markboth{
  B. Sadigh, P. Erhart, A. Stukowski, and A. Caro  
}{
  Composition-Dependent Interatomic Potentials
}

\maketitle

\begin{abstract}
We propose a simple scheme to construct composition-dependent
interatomic potentials for multicomponent systems that when
superposed onto the potentials for the pure elements can
reproduce not only the heat of mixing of the solid solution in
the entire concentration range but also the energetics of a wider
range of configurations including intermetallic phases.
We show that an expansion in cluster interactions provides a way to
systematically increase the accuracy of the model, and that it is
straightforward to generalise this procedure to multicomponent
systems.
Concentration-dependent interatomic potentials can be built upon
almost any type of potential for the pure elements including embedded
atom method (EAM), modified EAM, bond-order, and
Stillinger-Weber type potentials.
In general, composition-dependent $N$-body terms in the total energy
lead to explicit $N+1$-body forces, which potentially renders them
computationally expensive. We present an algorithm that overcomes this
problem and that can speed up the calculation of the forces for
composition-dependent pair potentials in such a way as to make them
computationally comparable in efficiency and scaling behaviour to
standard EAM potentials. We also discuss the implementation in
Monte-Carlo simulations.
Finally, we exemplarily review the composition-dependent EAM model for
the Fe--Cr system [PRL {\bf 95},075702, (2005)].
\end{abstract}


\section{Introduction}

Twenty-five years ago, the Finnis-Sinclair many body potential
\cite{FinSin84}, the Embedded Atom Model of Daw and Baskes
\cite{DawBas84}, the Glue model of Ercolessi and Parrinello
\cite{ErcParTos86}, and the effective medium theory due to Puska, Nieminen
and N\/{o}rskov \cite{PusNieMan81, Nor82} marked the birthday of
modern atomic scale computational materials science, enabling computer
simulations at the multimillion atom scale to become a routine in
modern materials science research. This family of many body potentials
share in common the fact that the expression for the total energy has
non linear contributions of pair functions, removing in this way the
limitations of the pair potential formulation to describe realistic
elastic properties.

Alloys and compounds, where the thermodynamic information is of
relevance, is one of the main fields in which these potentials have
been applied. In the early days of many body potentials the main alloy
property fitted was the heat of solution of a single impurity
\cite{FoiBasDaw86}, {\it i.e.} the dilute limit of the heat of
formation (HOF) of the alloy. However, when these potentials are
applied to concentrated alloys the predictions are usually
uncontrolled; they work well for systems with a mixing enthalpy that
is nearly symmetric and positive over the entire concentration range,
as for example in the cases of Fe--Cu \cite{LudFarPed98, PasMal07}, or
Au--Ni \cite{FoiBasDaw86, AstFoi96, ArrCarCar02}.

Alloys which show a strong asymmetry or even a sign inversion in the
HOF such as Fe--Cr or Pd--Ni are beyond the scope of standard many
body potential models, and there is not yet a unique methodology
suitable for their description. Similar limitations apply with respect
to systems with a negative HOF which feature intermetallic
phases. Frequently, such systems require different parametrisations
for different phases, as in the case of Ni--Al with the B2 phase on
one hand \cite{MisMehPap02}, and the $\gamma$ and $\gamma'$ phases on
the other \cite{Mis04}.

Two schemes have been developed to deal with these shortcomings in the
case of Fe--Cr which displays an inversion in the HOF as a function of
concentration, namely the composition-dependent embedded atom method
(CD-EAM) \cite{CarCroCar05} and the two-band model (2BM)
\cite{OlsWalDom05}. For neither one of these schemes, it is obvious how
it can be extended to more than two components.

The objective of this paper is to develop a framework for constructing
interatomic potential models for multicomponent alloys based on an
expansion in clusters of increasing size that can be practically
implemented and systematically improved. Our methodology allows to
describe systems with arbitrary heat of mixing curves and includes
intermetallic phases in a systematic and physically meaningful
fashion. Thereby, we overcome the most important disadvantages of
current alloy potential schemes and provide a framework for systems of
arbitrary complexity.

In our methodology the interatomic interactions are modified by
composition-dependent functions. This introduces a dependence on the
environment which is somewhat reminiscent of the bond-order potential
(BOP) scheme developed by Abell and Tersoff \cite{Abe85, Ter86,
  Ter88a}. In this formalism the attractive pair potential is
scaled by a (usually) angular dependent function (the ``bond-order'')
which describes the local structure. Thereby, it is possible to
distinguish different lattice structures (face-centred cubic,
body-centred cubic, cubic diamond {\it etc.}) and also to stabilise
structures with low packing density such as diamond or zincblende
lattices.
(In fact, the BOP formalism has been successfully applied to model
alloys such as Fe--Pt that feature intermetallic phases with different
lattice structures \cite{MulErhAlb07b}).
The composition-dependent interatomic potential (CDIP)
scheme introduced in the present work and the BOP formalism thus both
include explicit environment-dependent terms. However, in the CDIP
approach this environment-dependence is used to distinguish different {\it
  chemical} motifs while in the BOP scheme it is used to identify
different {\it structural} motifs.

This paper is organised as follows: In \sect{sect:pair_potentials} we
introduce the basic terminology and present a systematic approach to
fitting potentials for binary
systems. Section~\ref{sect:beyond_pair_potentials} describes how by
including higher order terms it is possible to fit e.g., intermetallic
phases. In \sect{sect:series} a series expansion is developed which
generalises the concepts introduced in the previous sections and which
is used in \sect{sect:ternary} to obtain explicit expressions for a
ternary system. The efficient computation of forces is discussed in
\sect{sect:forces} and an optimal implementation in Monte-Carlo
simulations is the subject of \sect{sect:monte_carlo}. Finally, as an
example, the composition-dependent embedded atom method potential
for Fe--Cr is reviewed in \sect{sect:FeCr}.

\section{Binary Systems}

\subsection{Pair Potentials}
\label{sect:pair_potentials}

For the sake of clarity of the following exposition, we assume EAM
models throughout this paper. It is important to stress that the
formalism to be developed hereafter can be applied to any potential
model for the pure elements including modified embedded atom method
(MEAM) \cite{Bas87, Bas92}, bond-order \cite{Abe85, Ter86, Ter88a},
and Stillinger-Weber type \cite{StiWeb85} potentials.

Consider a single-component system of atoms {\it A}, whose interactions are
described by the EAM model,
\begin{eqnarray}
  \label{eq:1}
  E_A = \sum_i U_A\left(\overline{\rho}_i \right)
  + \frac{1}{2} \sum_i\sum_j\phi_A\left(r_{ij}\right)
  \quad\text{with}\quad
  \overline{\rho}_i = \sum_{j\neq i} \rho(r_{ij}) .
  \label{eq:rho}
\end{eqnarray}
The first term in \eq{eq:1} contains the embedding function
$U_A(\overline{\rho}_i)$, which is a nonlinear function of the local
electron density $\overline{\rho}_i$ around atom $i$. It accounts for
cohesion due to band formation in the solid state and is constructed
to reproduce the equation of state of system {\it A}. The second term
represents the remainder of the interaction energy. It can be
interpreted as the effective screened Coulomb interaction between
pairs of ions in {\it A}. The EAM formalism can capture the energetics
associated with density fluctuations in the lattice and has been
successfully applied for modelling the formation of crystal defects
such as vacancies, interstitials and their clusters.

Consider now a binary system, where the pure phases are described by
EAM potentials.
It can be shown that the total energy expression for this type of
potentials is invariant under certain scaling operations
\cite{DawFoiBas93}. This ``effective pair format'' can be used to
rescale the two EAM potentials, e.g. such that at the equilibrium
volume for a certain lattice the electron density is 1, to ensure
their compatibility.
One part of the total energy of the two-component system can be
written as the superposition of the respective embedding terms and
effective pair interactions:
\begin{eqnarray}
  \label{eq:2}
  E_0 &=& \sum_{i\in A} U_A
  \left(\overline{\rho}_i^{A}+\mu_{A(B)}~\overline{\rho}_i^{B}\right) + 
  \frac{1}{2} \sum_{i\in A}\sum_{j\in A}\phi_A\left(r_{ij}\right)\\
  &+&
  \sum_{i\in B} U_B
  \left(\overline{\rho}_i^{B}+\mu_{B(A)}~\overline{\rho}_i^{A}\right) + 
  \frac{1}{2} \sum_{i\in B}\sum_{j\in B}\phi_B\left(r_{ij}\right) ,\nonumber
\end{eqnarray}
where
\begin{equation}
  \label{eq:rhobar}
  \overline{\rho}_i^{\mathcal{S}}
  = \sum_{j\in \mathcal{S},j\neq i} \rho^{\mathcal{S}}(r_{ij}).
\end{equation}
Note that above we have not yet added any explicit $A-B$
interactions. Equation~(\ref{eq:2}) is a strict superposition of the
interatomic potentials for the pure elements with the only caveat that
the electron density of the $A$ ($B$) species in the embedding
function of a $B$ ($A$) particle is scaled with a parameter
$\mu_{B(A)}$ in order to account for the different local electron
densities.
Thereby, two EAM models can be calibrated with respect to each
other. More elaborate schemes are possible, e.g. one can treat $\mu_A$
and $\mu_B$ as free parameters. Here for the sake of simplicity, we
restrict ourselves to normalised electron densities.

Starting from a parametrisation for $E_0$, we now devise a practical
scheme for systematically improving the interaction model. Let us
denote the true many-body energy functional of the binary system by
$E_t$. Our goal is to construct an interatomic potential model for the
difference energy functional $\Delta E^{(0)} = E_t-E_0$. We begin with
the two dilute limits. Consider a lattice of $A$ particles and
substitute the atom residing in the $i$-th site with a $B$ atom. Let
us now assume that $\Delta E^{(0)}$ for this configuration can be
satisfactorily represented by a pair potential between the $A-B$
pairs. In this limit $\Delta E^{(0)}$ can thus be written as
\begin{equation}
  \label{eq:4}
  \Delta E^{(0)}({\text{$A$-rich}}) = \sum_{j \in A} V^{A}_{AB}(r_{ij}).
\end{equation} 
(There is only one sum in this expression since we are dealing with a
configuration that contains only one $B$ atom). A similar expression
is obtained for the $B$-rich limit
\begin{equation}
  \label{eq:5}
  \Delta E^{(0)}({\text{$B$-rich}}) = \sum_{j \in B} V^{B}_{AB}(r_{ij}). 
\end{equation}
Since we do not require the pair potential models for the two dilute
limits to coincide with each other, an interpolation is needed which
preserves the energetics of the impurities. The main objective of the
present paper is to devise such an interpolation scheme. The simplest
ansatz for such an expression is
\begin{equation}
\label{eq:dE}
\Delta E^{(0)} = \sum_{i \in A}\sum_{j\in B} x^A_{ij} V^A_{AB}(r_{ij}) + 
\sum_{i \in A}\sum_{j\in B} x^B_{ij} V^B_{AB}(r_{ij}) 
\end{equation}
Above, $x^{\mathcal{S}}_{ij}$ denotes the concentration of species
$\mathcal{S}$ in the neighbourhood of an $A-B$ pair residing on the $i$
and $j$ sites. Ideally, we require this quantity to be easy to
calculate and to be insensitive to the local density and topology, in
other words it should separate chemistry from structure. In any case,
$x^{\mathcal{S}}_{ij}$ has to represent an average over the
neighbourhood of both centres $i$ and $j$. Before we derive the
expression for $x^{\mathcal{S}}_{ij}$, it is instructive to discuss
the corresponding one-centre quantity $x^{\mathcal{S}}_i$. It
describes the local concentration of species $\mathcal{S}$ around atom
$i$. A simple way to determine $x^{\mathcal{S}}_i$ is to choose a
local density function $\sigma(r_{ij})$ and then to evaluate the
following expression
\begin{equation}
  x^{\mathcal{S}}_i
  = \frac{\sum_{(j\in S,j\neq i)}\sigma(r_{ij})}{\sum_{j\neq i}\sigma(r_{ij})}
  = \frac{\overline{\sigma}^{\mathcal{S}}_i}{\overline{\sigma}_i},
  \label{eq:conc_i}
\end{equation}
which is indeed rather insensitive to the local geometry. This is most
obvious in the dilute limits. The local concentration
$x^{\mathcal{S}}_i$ at the site of an impurity atom $i$ is either 0
(if $\mathcal{S}$ is the minority species) or 1 (if $\mathcal{S}$ is
the majority species) independent of the local structure. This is,
however, strictly true only for the impurity atom. For the other atoms
in the system $x^{\mathcal{S}}_j$ varies between 0 and 1 depending on
the distance to the impurity atom. Also for these particles, atomic
displacements may change the value of $x^{\mathcal{S}}_j$.  A total
decoupling of chemistry and structure is therefore not possible. The
optimal choice for $\sigma(r_{ij})$ is the function that minimises the
effect of local geometry on $x^{\mathcal{S}}_i$. Although it is
possible to choose different $\sigma$-functions for the different
species, we do not expect the quality of the final potential to depend
crucially on the choice of $\sigma(r_{ij})$. In fact, we expect the
best choice for $\sigma(r_{ij})$ to be the simplest one.

\begin{figure}
  \centering
\includegraphics[scale=0.11]{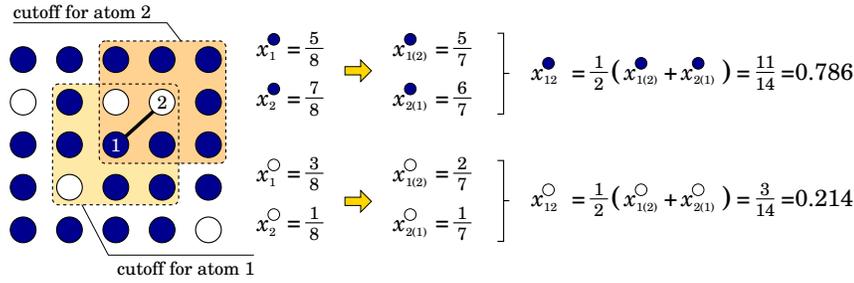}
  \caption{
    Schematic illustration of the connection between
    $x_i^{\mathcal{S}}$ and two-centre concentrations
    $x_{ij}^{\mathcal{S}}$ and their computation in a binary alloy
    according to Eqs.~(\ref{eq:conc_i}) and (\ref{eq:conc_ij}).
    Here, the cutoff function $\sigma(r)$ which appears in
    \eq{eq:conc_i} is assumed to be a step function which is 1 for
    $r<r_c$ and zero otherwise.
  }
  \label{fig:schematic_2comp}
\end{figure}

It is now straightforward to extend \eq{eq:conc_i} to define the
concentration $x^{\mathcal{S}}_{ij}$ in the neighbourhood of a pair of
atoms residing on sites $i$ and $j$. To this end, we first define a
quantity $x^{\mathcal{S}}_{i(j)}$ to represent the concentration of
the species $\mathcal{S}$ in the neighbourhood of atom $i$ excluding
atom $j$:
\begin{eqnarray}
\label{eq:conc_ij}
  x^{\mathcal{S}}_{i(j)}
  &=& \frac{\sum_{(k\in \mathcal{S},k\neq i,k\neq j)}\sigma(r_{ik})}{
    \sum_{(k\neq i,k\neq j)}\sigma(r_{ik})}
  =
  \frac{\overline{\sigma}^{\mathcal{S}}_i 
    - \delta(\mathcal{S},t_j) \sigma(r_{ij})}{
    \overline{\sigma}_i - \sigma(r_{ij})}
  \\
  &=&
  x^{\mathcal{S}}_i
  \left\{
  \begin{array}{ll}
    \displaystyle
    \frac{1-\sigma(r_{ij})/\overline{\sigma}^{\mathcal{S}}_i}{
      1-\sigma(r_{ij})/\overline{\sigma}_i}
    & ~ t_j = \mathcal{S}
    \\
    \displaystyle
    \frac{1}{1-\sigma(r_{ij})/\overline{\sigma}_i}
    & ~ t_j = \mathcal{S}
  \end{array}
  \right.
  ,
  \nonumber
\end{eqnarray}
where $t_i$ denotes the type of atom $i$, and $\delta(t_i,t_j)$ is 1
if $t_i=t_j$ and zero otherwise. Using this quantity, the two-centre
concentration $x^{\mathcal{S}}_{ij}$ can be defined as follows
\begin{equation}
  x^{\mathcal{S}}_{ij}
  = \frac{1}{2}\left(x^{\mathcal{S}}_{i(j)} + x^{\mathcal{S}}_{j(i)} \right)
  \label{eq:2cntr}
\end{equation}
Hence, the two-centre concentration of the species $\mathcal{S}$ about
the atom pair $(i,j)$ is the average concentration in the two separate
neighbourhoods of sites $i$ and $j$ excluding both of these atoms. This
definition, which is illustrated in \fig{fig:schematic_2comp}, has the
important advantage that the interpolation scheme introduced in
\eq{eq:dE} does not modify the interactions in the dilute limits,
since $x^{\mathcal{S}}_{ij}$ is strictly 0 or 1 in the two limits
irrespective of the local structure. Furthermore, it is
straightforward to generalise \eq{eq:2cntr} to multi-centre
concentrations. For example, in the next section, we will explicitly
discuss the construction of interatomic potentials using three-centre
concentrations.

Let us now revisit \eq{eq:dE}. As mentioned earlier this is the
simplest ansatz for $\Delta E^{(0)}$ that can reproduce the energetics
of both dilute limits. A more general expression is
\begin{equation}
  \label{eq:dE1}
  \Delta E^{(0)} = \sum_{i \in A}\sum_{j\in B} h_{AB}^A(x^A_{ij})~ V^A_{AB}(r_{ij}) + 
  \sum_{i \in A}\sum_{j\in B} h_{AB}^B(x^B_{ij})~V^B_{AB}(r_{ij}) ,
\end{equation}
where $h_A^B(x)$ and $h_B^A$ are nonlinear functions with the property
$h_A^B(0) = h_B^A(0) = 0$ and $h_A^B(1) = h_B^A(1) = 1$. By fitting
these functions to the energetics of the concentrated alloys, the
quality of the interatomic potential model for the binary can be
improved drastically.

In principle, one can stop here and have an interatomic potential
model, $E_0 + \Delta E^{(0)}$, that can reproduce the energetics of
the dilute limits as well as the solid solution of the binary. It is,
however, also possible to further refine the above model. For this
purpose, let us again define a difference energy functional
\begin{equation}
  \Delta E^{(1)} = E_t - E_0 - \Delta E^{(0)},
\end{equation}
and construct an interatomic potential model for the energy functional
$\Delta E^{(1)}$. Consider a lattice of $A$ particles and substitute
two atoms, say $i$ and $j$, with $B$ particles. Assume that $\Delta
E^{(1)}$ for this configuration can be well represented by a potential
model describing the interaction of the $B$-$B$ pair with a lattice of $A$
particles. In this limit we can express $\Delta E^{(1)}$ as
\begin{equation}
  \label{eq:Arich1}
  \Delta E^{(1)}({\text{$A$-rich}}) = V_{BB}^A(r_{ij})+\sum_k V_{BBA}^A(r_{ijk}),
\end{equation}
where $r_{ijk}$ is shorthand for the three sets of positions of the
$i$, $j$ and $k$ atoms,
i.e. $\{\vec{r}_{i},\vec{r}_{j},\vec{r}_{k}\}$. In the same way we
obtain for the $B$-rich limit
\begin{equation}
  \label{eq:Brich1}
  \Delta E^{(1)}({\text{$B$-rich}}) = V_{AA}^B(r_{ij})+\sum_k V_{AAB}^B(r_{ijk}).
\end{equation}
Note that $\Delta E^{(1)}$ has both a two-body and a three-body
component and thus can be decomposed as follows
\begin{equation}
  \label{eq:15}
  \Delta E^{(1)} = \Delta E^{(1)}_{\text {pair}} + \Delta E^{(1)}_{\text {triplet}} .
\end{equation}
In the next section we discuss how to incorporate the three-body
contribution into the interatomic potential model. For now, we only
consider $\Delta E^{(1)}_{\text {pair}}$. Following the same line of
arguments that lead to \eq{eq:dE1}, we obtain the expression
\begin{equation}
  \Delta E^{(1)}_{\text {pair}} = \sum_{i \in B}\sum_{j\in B} h_{BB}^A(x^A_{ij})~ V^A_{BB}(r_{ij}) + 
  \sum_{i \in A}\sum_{j\in A} h_{AA}^B(x^B_{ij})~V^B_{AA}(r_{ij}) ,
\end{equation}
which reproduces the contributions of the pair terms in the two limits
given by Eqs.~(\ref{eq:Arich1}) and (\ref{eq:Brich1}). The two
non-linear functions have to fulfil the conditions
\begin{eqnarray}
h_{AA}^B(0) = h_{BB}^A(0) = 0 \\
h_{AA}^B(1) = h_{BB}^A(1) = 1.
\end{eqnarray}
By fitting the functions $h_{AA}^B$ and $h_{BB}^A$ in the intermediate
concentration range to the energetics of the concentrated alloy, one
can obtain a further improvement for the interaction model for the
binary system.

\subsection{Beyond Pair Potentials}
\label{sect:beyond_pair_potentials}

In this section, we show that the formalism introduced in the previous
section can be extended to multi-body interaction potentials, which
enables us to capture the energetics of a wider range of phases
including ordered compounds. In the previous section, we outlined a
scheme to construct composition-dependent pair potentials for the
potential energy landscape $E_0 + \Delta E^{(0)} + \Delta E^{(1)}$. It
was also observed that a proper formulation of $\Delta E^{(1)}$
requires incorporation of explicit three-body terms. In this section
we describe how to construct such composition-dependent multi-body
potentials.

\begin{figure}
  \centering
\includegraphics[scale=0.11]{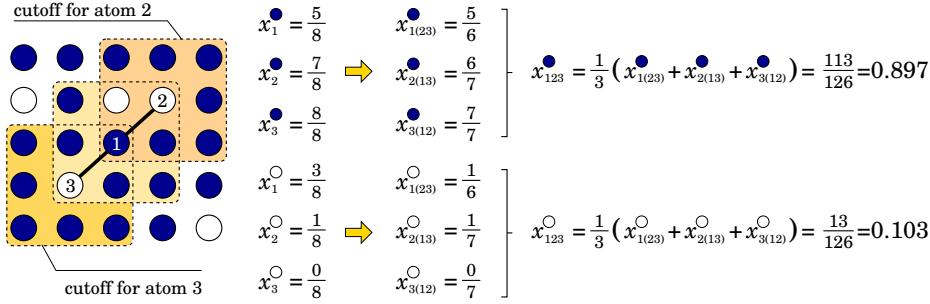}
  \caption{
    Schematic illustration of the computation of three-centre
    concentrations in a binary alloy using Eqs.~(\ref{eq:conc_i}) and
    (\ref{eq:conc_ijk}).
    Here, the cutoff function $\sigma(r)$ which appears in
    \eq{eq:conc_i} is assumed to be a step function which is 1 for
    $r<r_c$ and zero otherwise.
  }
  \label{fig:schematic_2comp_trip}
\end{figure}

First, we require an interpolation scheme to connect the two limits of
the three-body term $\Delta E^{(1)}_{\text {triplet}}$ in
\eq{eq:15}. The simplest ansatz for such an expression is
\begin{equation}
  \label{eq:triplet}
  \Delta E^{(1)}_{\text {triplet}} = 
  \sum_{i \in B}\sum_{j\in B}\sum_{k\in A} x^A_{ijk} V^A_{BBA}(r_{ijk}) + 
  \sum_{i \in A}\sum_{j\in A}\sum_{k\in B} x^B_{ijk} V^B_{AAB}(r_{ijk}) ,
\end{equation}
where $x^{\mathcal {S}}_{ijk}$ denotes the concentration of species
$\mathcal {S}$ in the neighbourhood of the triplet residing on sites
$i$, $j$ and $k$. In analogy with the derivation of the two-centre
concentration \eq{eq:2cntr}, we start from the one-centre
concentration $x^{\mathcal {S}}_i$ and define the intermediate
quantity $x^{\mathcal{S}}_i{(jk)}$ that represents the concentration
centred around atom $i$ excluding atoms $j$ and $k$
\begin{eqnarray}
  x^{\mathcal{S}}_{i(jk)}
  &=&
  \frac{\sum_{(l\in S,l\neq i,l\neq j,l\neq k)}\sigma(r_{il})}
       {\sum_{(l\neq i,l\neq j,l\neq k)}\sigma(r_{il}) } 
  = \frac{\overline{\sigma}^{\mathcal{S}}_i
    - \delta(\mathcal{S},t_j) \sigma(r_{ij})
    - \delta(\mathcal{S},t_k) \sigma(r_{ik})}{
    \overline{\sigma}_i - \sigma(r_{ij})-\sigma(r_{ik})}  \\
  &=&
  x^{\mathcal{S}}_i~\frac{1
    -\left[ \delta(\mathcal{S},t_j) \sigma(r_{ij})
    +       \delta(\mathcal{S},t_k) \sigma(r_{ik}) \right]
    /\overline{\sigma}^{\mathcal{S}}_i}{
    1 - \left[\sigma(r_{ij}) + \sigma(r_{ik})\right]/\overline{\sigma}_i}, 
\end{eqnarray}
and now following the same line of arguments leading to \eq{eq:2cntr}
we define the three-centre concentration $x^{\mathcal{S}}_{ijk}$ as
follows
\begin{equation}
  x^{\mathcal{S}}_{ijk}
  = \frac{1}{3}\left(x^{\mathcal{S}}_{i(jk)}
  + x^{\mathcal{S}}_{j(ik)} + x^{\mathcal{S}}_{k(ij)}\right).
  \label{eq:conc_ijk}
\end{equation}
A graphical illustration of the computation of this quantity is given
in \fig{fig:schematic_2comp_trip}.
The three-centre concentration of the species $\mathcal{S}$ about the
triplet $(i,j,k)$ is the average concentration (excluding the triplet)
in three separate neighbourhoods, each of which is centred at one of
the atoms in the triplet. Thanks to this definition
$x^{\mathcal{S}}_{ijk}$ is strictly 0 or 1 in the two dilute limits
described in Eqs.~(\ref{eq:Arich1}) and (\ref{eq:Brich1}),
irrespective of the local structure. Hence, the interpolation scheme
in \eq{eq:triplet} does not alter the interactions in
Eqs.~(\ref{eq:Arich1}) and (\ref{eq:Brich1}). Again, as in
\eq{eq:dE1}, we can improve the simple interpolation scheme in
\eq{eq:triplet}
\begin{equation}
  \label{eq:triplet1}
  \Delta E^{(1)}_{\text {triplet}} = 
  \sum_{i \in B}\sum_{j\in B}\sum_{k\in A} h_{BBA}^A(x^A_{ijk}) V^A_{BBA}(r_{ijk}) + 
  \sum_{i \in A}\sum_{j\in A}\sum_{k\in B} h_{AAB}^B(x^B_{ijk}) V^B_{AAB}(r_{ijk}) ,
\end{equation}
where $h_{BBA}^A$ and $h_{AAB}^B$ are non-linear functions that can be
fitted to the energetics of the concentrated alloys with the boundary
conditions
\begin{eqnarray}
  h_{BBA}^A(0) = h_{AAB}^B(0) = 0
  \quad\text{and}\quad
  h_{BBA}^A(1) = h_{AAB}^B(1) = 1.
\end{eqnarray}
Following this scheme composition-dependent cluster interactions of
arbitrary order can be included in the interatomic potential model. To
summarise, to incorporate cluster interactions of order $n$, two
cluster potentials are constructed, one for the configuration where
the cluster is embedded in the $A$ lattice and one for the
configuration where the cluster is embedded in the $B$
lattice. Subsequently these limits are interpolated using the
$n$-centre concentrations. In the next section, we review this
strategy in detail to show that a systematic series expansion in
composition-dependent cluster interactions is possible for general
multicomponent systems.

\section{Multicomponent Systems}

\subsection{Series Expansion in Embedded Cluster Interactions}
\label{sect:series}

In the first sections of this paper, we have shown how to practically
construct interatomic potentials for binary systems. First, mixed
interatomic pair and triplet potentials are generated for the dilute
limits which are subsequently extended to arbitrary concentrations by
fitting interpolation functions that depend on the local concentration
about the atomic pairs and triplets. The choice of specific potentials
and dilute configurations was mainly driven by physical intuition. In
this section we show that this procedure can be formalised and
generalised to arbitrarily complex systems with more than two
components.

\begin{figure}
  \centering
\includegraphics[scale=0.11]{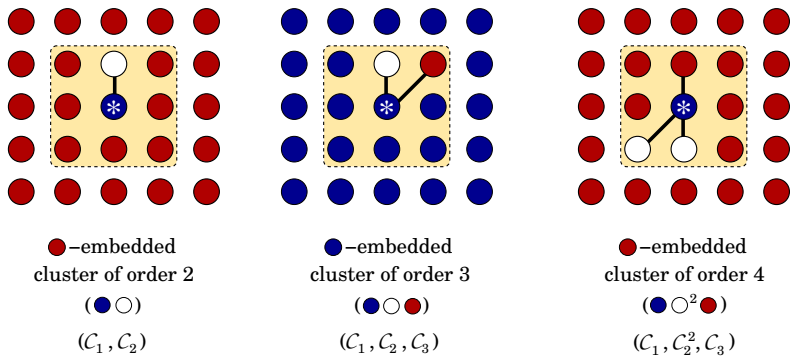}
  \caption{
    Schematic illustration of $\mathcal{S}$-embedded coloured clusters
    of orders 2, 3, and 4 in a ternary alloy. The shaded region
    indicates the cutoff range around the central atom marked by an
    asterisk.
  }
  \label{fig:schematic_clusters1}
\end{figure}

Consider an $n$-component mixture of $N$ particles that are
distinguishable only through their species. Assign a unique colour to
each of the species: $\{\mathcal{C}_1,\ldots,\mathcal{C}_n\}$. We
define a colour cluster of order $m$ to be a set of $m$ particles with
a specific colour combination. We use the occupation number formalism
to identify colour schemes,
i.e. $\left(\mathcal{C}_1^{k_1},\ldots,\mathcal{C}_n^{k_n}\right)$,
where $k_i$ is the number of particles in the cluster with colour
$\mathcal{C}_i$, and $\sum_i k_i = m$. For example, a cluster of order
3 consisting of one particle with the colour $\mathcal{C}_1$ and two
particles with the colour $\mathcal{C}_3$, is denoted by
$\left(\mathcal{C}_1,\mathcal{C}_3^2\right)$. Furthermore, we define
an $\mathcal{S}$-embedded colour cluster of order $m$ to be a set of
$m$ coloured particles embedded in a pure matrix of species
$\mathcal{S}$. Three examples of such $\mathcal{S}$-embedded coloured
clusters are shown in \fig{fig:schematic_clusters1}. The key idea is
that the potential energy landscape of an alloy can be expanded in the
basis set of elementary interaction potentials each of which is
constructed to reproduce the energetics of a particular embedded colour
cluster. The order of an interaction element in the series is
determined by the order of the corresponding colour cluster. By
progressively including higher order colour cluster interactions, one
can systematically increase the accuracy of the model.

\begin{figure}
  \centering
\includegraphics[scale=0.11]{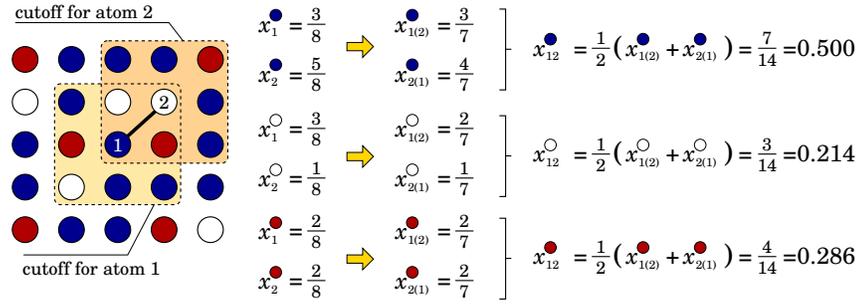}
  \caption{
    Schematic illustration of the connection between
    $x_i^{\mathcal{S}}$ and two-centre concentrations
    $x_{ij}^{\mathcal{S}}$ and their computation in a ternary alloy
    according to Eqs.~(\ref{eq:conc_i}) and (\ref{eq:conc_ij}).
    Here, the cutoff function $\sigma(r)$ which appears in
    \eq{eq:conc_i} is assumed to be a step function which is 1 for
    $r<r_c$ and zero otherwise.
  }
  \label{fig:schematic_3comp}
\end{figure}

To recapitulate, we expand the potential energy landscape of
multicomponent systems in the basis set of colour cluster interatomic
potential functions $V^{\mathcal{S}}_{
  \mathcal{C}_1^{k_1}\ldots\mathcal{C}_n^{k_n} } (\{ \vec{r} \} )$, where
$\{ \vec{r} \}$ is the real-space configuration of the respective
cluster. The expansion coefficient for each basis function is the
interpolation function $h^{\mathcal{S}}_{
  \mathcal{C}_1^{k_1}\ldots\mathcal{C}_n^{k_n}} (x^{\mathcal{S}})$,
where $x^{\mathcal{S}}$ is the local concentration of the species
$\mathcal{S}$ in the neighbourhood of the cluster. One of the
innovations in this work is a simple and computationally expeditious
way to determine $x^{\mathcal{S}}$ which is illustrated for the case
of a ternary alloy in \fig{fig:schematic_3comp}. Formally the total
energy expression for an alloy of $n$ components and $N$ particles can
be written as
\begin{equation}
  E = E_0
  + \sum_m \underbrace{\sum_{k_1}\ldots\sum_{k_n}}_{\sum_{i=1}^n k_i=m}\sum_{\mathcal{S}}
  h^{\mathcal{S}}_{ \mathcal{C}_1^{k_1}\ldots\mathcal{C}_n^{k_n}} (x^{\mathcal{S}})
  V^{\mathcal{S}}_{ \mathcal{C}_1^{k_1}\ldots\mathcal{C}_n^{k_n} } (\{ \vec{r} \} ),
\end{equation}
where the first sum is over the order of the cluster potentials and
the subsequent sums are over all distinguishable colour combinations of
$m$-size clusters. Each term in the above expansion can be evaluated
as follows
\begin{equation}
  \label{eq:27}
  h^{\mathcal{S}}_{ \mathcal{C}_1^{k_1}\ldots\mathcal{C}_n^{k_n}} (x^{\mathcal{S}})
  V^{\mathcal{S}}_{ \mathcal{C}_1^{k_1}\ldots\mathcal{C}_n^{k_n} } (\{ \vec{r} \} )
  =
  \underbrace{\sum_{i_1=1}^N\ldots\sum_{i_m=1}^N }_{
    \{i_1\ldots i_m\}\in \{ \mathcal{C}_1^{k_1}\ldots\mathcal{C}_n^{k_n} \} }
  h^{\mathcal{S}}_{ \mathcal{C}_1^{k_1}\ldots\mathcal{C}_n^{k_n}} (x^{\mathcal{S}}_{i_1\ldots i_m}) 
  V^{\mathcal{S}}_{ \mathcal{C}_1^{k_1}\ldots\mathcal{C}_n^{k_n} } (r_{i_1\ldots i_m}).
\end{equation}
The sums in \eq{eq:27} are over all possible $m$-size atom clusters
$\{i_1\ldots i_m\}$ in the system with the colour scheme
$\left(\mathcal{C}_1^{k_1},\ldots,\mathcal{C}_n^{k_n}\right)$.

The main advantage of this scheme is that the basis functions can be
constructed sequentially and independent of the interpolation
functions. The lower order terms can be constructed with no knowledge
of the higher order terms and therefore need not be reparametrised
when higher order cluster potentials are constructed. The higher order
terms in the expansion become progressively smaller. 
Furthermore, addition of new terms in the series expansion is not
likely to introduce unphysical behaviour, a problem that plagues most
fitting schemes for interatomic potentials.

\subsection{Explicit expressions for ternary alloys}
\label{sect:ternary}

In this section we illustrate the formal discussion in the previous
section by constructing an expansion in embedded pair and triplet
potentials for a ternary system. For simplicity we assume the pure
elements are described by EAM models. The extension to larger number
of components and higher order cluster potentials will be obvious. We
consider a system of three components $A$, $B$ and $C$, and assume
that three composition-dependent pair potentials for the binary
systems $A-B$, $A-C$ and $B-C$ have already been
constructed. Explicitly, the $A-B$ interaction is given by the
following expression
\begin{eqnarray}
  E^{pair}_{\text {A-B}} &=& \sum_{i\in A} U_A
  \left(\overline{\rho}_i^{A}+\mu_{A(B)}~\overline{\rho}_i^{B}\right) + 
  \frac{1}{2} \sum_{i\in A}\sum_{j\in A}\left( h_{AA}^A(x^A_{ij}) \phi_A(r_{ij}) + 
  h_{AA}^B(x^B_{ij}) V_{AA}^B(r_{ij})  \right) \nonumber\\
  &+&  \sum_{i\in B} U_B\left(\overline{\rho}_i^{B}+\mu_{B(A)}~\overline{\rho}_i^{A}\right) + 
  \frac{1}{2} \sum_{i\in B}\sum_{j\in B}\left(h_{BB}^B(x^B_{ij}) \phi_B(r_{ij}) + 
  h_{BB}^A(x^A_{ij}) V_{BB}^A(r_{ij})  \right)  \nonumber\\ 
  &+& \sum_{i\in A}\sum_{j\in B}\left( h_{AB}^A(x^A_{ij}) V_{AB}^A(r_{ij}) + 
  h_{AB}^B(x^B_{ij}) V_{AB}^B(r_{ij}) \right).
\end{eqnarray} 
By now the notation above should be familiar. The interaction
potentials for the two other pairs can be written in analogous
fashion.

Now, we can spell out the expansion in embedded pair potentials for
the ternary $A-B-C$
\begin{eqnarray}
  \label{eq:ternary}
  E^{pair}_{A-B-C} &=&
  \sum_{i\in A} U_A\left(\overline{\rho}_i^{A}+\mu_{A(B)}~\overline{\rho}_i^{B} +
  \mu_{A(C)}~\overline{\rho}_i^{C}\right) \\ 
  &+&  \sum_{i\in B} U_B\left(\overline{\rho}_i^{B}+\mu_{B(A)}~\overline{\rho}_i^{A} 
  + \mu_{B(C)}~\overline{\rho}_i^{C}\right)  \nonumber\\
  &+&  \sum_{i\in C} U_C\left(\overline{\rho}_i^{C}+\mu_{C(A)}~\overline{\rho}_i^{A}+ 
  \mu_{C(B)}~\overline{\rho}_i^{B}\right) \nonumber \\
  &+&\frac{1}{2} \sum_{i\in A}\sum_{j\in A}\left[ h_{AA}^A(x^A_{ij}) \phi_A(r_{ij}) + 
  h_{AA}^B(x^B_{ij}) V_{AA}^B(r_{ij}) + h_{AA}^C(x^C_{ij}) V_{AA}^C(r_{ij})\right] \nonumber\\
  &+&\frac{1}{2} \sum_{i\in B}\sum_{j\in B}\left[ h_{BB}^B(x^B_{ij}) \phi_B(r_{ij}) + 
  h_{BB}^A(x^A_{ij}) V_{BB}^A(r_{ij}) + h_{BB}^C(x^C_{ij}) V_{BB}^C(r_{ij})\right] \nonumber\\
  &+&\frac{1}{2} \sum_{i\in C}\sum_{j\in C}\left[ h_{CC}^C(x^C_{ij}) \phi_C(r_{ij}) + 
  h_{CC}^A(x^A_{ij}) V_{CC}^A(r_{ij}) + h_{CC}^B(x^B_{ij}) V_{CC}^B(r_{ij})\right] \nonumber\\
  &+& \sum_{i\in A}\sum_{j\in B}\left[ h_{AB}^A(x^A_{ij}) V_{AB}^A(r_{ij}) + 
  h_{AB}^B(x^B_{ij}) V_{AB}^B(r_{ij})  + h_{AB}^C(x^C_{ij}) V_{AB}^C(r_{ij}) \right] \nonumber\\
  &+& \sum_{i\in A}\sum_{j\in C}\left[ h_{AC}^A(x^A_{ij}) V_{AC}^A(r_{ij}) + 
  h_{AC}^B(x^B_{ij}) V_{AC}^B(r_{ij})  + h_{AC}^C(x^C_{ij}) V_{AC}^C(r_{ij}) \right] \nonumber\\
  &+& \sum_{i\in B}\sum_{j\in C}\left[ h_{BC}^A(x^A_{ij}) V_{BC}^A(r_{ij}) + 
  h_{BC}^B(x^B_{ij}) V_{BC}^B(r_{ij})  + h_{BC}^C(x^C_{ij}) V_{BC}^C(r_{ij}) \right] \nonumber.
\end{eqnarray}
The only unknowns in the above equation are $V_{AB}^C(r_{ij})$,
$V_{AC}^B(r_{ij})$, $V_{BC}^A(r_{ij})$, $h_{AB}^C(x^{C}_{ij})$,
$h_{AC}^B(x^{B}_{ij})$  and $h_{BC}^A(x^{A}_{ij})$. The potentials
$V_{AB}^C(r_{ij})$, $V_{AC}^B(r_{ij})$ and $V_{BC}^A(r_{ij})$ describe
the interaction between pairs of unlike species embedded in pure
lattices of the third species of the ternary. In analogy with the
previous section, it is reasonable to expect that we can construct
these potentials separately in their respective dilute limits and
subsequently fit the interpolation functions $h_{AB}^C(x^{C}_{ij})$,
$h_{AC}^B(x^{B}_{ij})$, $h_{BC}^A(x^{A}_{ij})$ to the energetics of
the concentrated ternary alloys. However, when the number of species
increases certain complications can arise that are not present in the
binaries. This is well illustrated in the situation above. We now show
that it is in fact not possible to separately construct the three pair
potentials $V_{AB}^C(r_{ij})$, $V_{AC}^B(r_{ij})$ and
$V_{BC}^A(r_{ij})$ described above. 

\begin{figure}
  \centering
\includegraphics[scale=0.11]{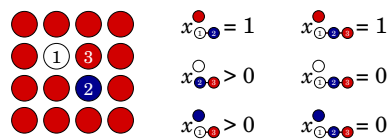}
  \caption{
    Schematic illustration of two and three-centre concentrations for
    a ternary alloy in the dilute limit. Note that the two-centre
    concentrations $x_{ij}$ in the {\it dilute limit} in a binary
    alloy are either one or zero. In contrast, in the case of a
    ternary alloy the two-centre concentrations in the same limit can
    be non-zero. The three-centre concentrations, however, are
    again either one or zero.
  }
  \label{fig:schematic_colored_cluster}
\end{figure}

To this end, consider a pure lattice of $N$ particles of e.g., $C$
species. Substitute two nearest neighbour particles in this lattice
with an $A$ particle and a $B$ particle respectively. The ternary
energy \eq{eq:ternary} for a $C$-rich configuration containing one
$A-B$ pair on the sites $i$ and $j$ respectively becomes
\begin{eqnarray}
  \label{eq:3dilute}
  E^{pair}_{A-B-C} ({\text{$C$-rich}}) &=& \tilde{E}_0\nonumber\\
  &+&\frac{1}{2} \sum_{k\in C}\sum_{l\in C}\left( h_{CC}^C(x^C_{kl}) \phi_C(r_{kl}) + 
  h_{CC}^A(x^A_{kl}) V_{CC}^A(r_{kl}) + h_{CC}^B(x^B_{kl}) V_{CC}^B(r_{kl})\right) \nonumber\\
  &+& V_{AB}^C(r_{ij})  + \sum_{k\in C}\left(
  h_{AC}^B(x^B_{ik}) V_{AC}^B(r_{ik})  + h_{AC}^C(x^C_{ik}) V_{AC}^C(r_{ik}) \right) \nonumber.\\
  &+& \sum_{k\in C} \left( h_{BC}^A(x^A_{jk}) V_{BC}^A(r_{jk}) + 
  h_{BC}^C(x^C_{jk}) V_{BC}^C(r_{jk}) \right) \nonumber,
\end{eqnarray}
where for the sake of clarity we have replaced the three embedding
terms in \eq{eq:ternary} by $\tilde{E}_0$. Observe that all three unknown
potentials $V_{AB}^C(r_{ij})$, $V_{AC}^B(r_{ik})$ and
$V_{BC}^A(r_{jk})$ as well as their corresponding interpolation
functions appear in \eq{eq:3dilute}. This is in contrast to the binary
case, e.g. Eqs.~(\ref{eq:4}), (\ref{eq:5}), (\ref{eq:Arich1}) and
(\ref{eq:Brich1}), where the potentials for the two dilute limits can
be constructed independently of each other. This is because the
two-centre concentrations in the {\it dilute limit} in a {\it binary}
alloy are either one or zero. In contrast, in the case of a {\it
  ternary} alloy the two-centre concentrations in the same limit can
be non-zero (see \fig{fig:schematic_colored_cluster}).

A straightforward solution to the above problem is to fit all three
pair potentials simultaneously. A closer look at \eq{eq:3dilute},
however, suggests a simpler solution. Let us examine the interpolation
functions $h_{AC}^B(x^B_{ik})$ and $h_{BC}^A(x^A_{jk})$. Note that
since we are dealing here with an $A-B$ cluster in a $C$-rich system
$x^B_{ik}$ and $x^A_{jk}$ are close to zero. Remembering the boundary
conditions on the interpolation functions, i.e. $h(1) = 1$ and $h(0) =
0$, we conclude that the contributions of the $V_{AC}^B(r_{ij})$ and
$V_{BC}^A(r_{ij})$ potentials to the energetics of an $A-B$ pair
embedded in a $C$ lattice are small. In fact, we can diminish the
contribution of these potentials to \eq{eq:3dilute} by enforcing the
interpolation functions to be 0 for $x < x_{th}$, where $x_{th}$ is
the largest concentration of $B$ or $A$ particles found about any pair
in the system. In this way, one can generally separate the
construction of cluster potentials when they overlap in the dilute
configurations.

The problem of potential overlap in the dilute limit discussed above
should not be neglected. On the other hand it is quite benign and
---as shown above--- can be handled easily. Furthermore, more often
than not, even for complex clusters and many components, there is no
overlap. We illustrate this point by considering the simplest
expansion in triplet cluster potentials for the ternary above:
\begin{eqnarray}
  E^{\text{triplet}}_{A-B-C} &=& \sum_{i\in A}\sum_{j\in B}\sum_{k\in C}
  h^A_{ABC}(x_{ijk}^A)~V^A_{ABC}(r_{ijk})  \\
  &+& h^B_{ABC}(x_{ijk}^B)~V^B_{ABC}(r_{ijk}) 
  + h^C_{ABC}(x_{ijk}^C)~V^C_{ABC}(r_{ijk}) \nonumber.
\end{eqnarray}
Now consider again the same $C$ lattice as above, where an $A-B$ pair
has been embedded at the sites $i$ and $j$. The triplet energy becomes
\begin{equation}
  E^{\text{triplet}}_{A-B-C}(\text{$C$-rich}) = \sum_{k\in C} V^C_{ABC}(r_{ijk}).
\end{equation}
Since we have only contributions from $V^C_{ABC}(r_{ijk})$ for the
these configurations, we can construct these potentials separately
from each other and independent of the interpolation functions.
This is because in the dilute limit the three-centre concentrations
are again either one or zero (see \fig{fig:schematic_colored_cluster}).

\section{Implementation of Forces in Molecular dynamics}
\label{sect:forces}

Next to accuracy, the most important quality of an interatomic
potential model is its computational efficiency when implemented into
atomistic simulation codes. Due to the unconventional form of the
interatomic potentials described in this work, it is important to
discuss the efficient implementation of forces for molecular-dynamics
simulations. We will see below that the straightforward derivation of
the forces for composition-dependent pair potentials leads to
explicit 3-body forces. In fact in general, composition-dependent
$N$-body potentials lead to explicit $N+1$-body forces. Below we
present an algorithm that considerably speeds up the calculation of
forces for composition-dependent $N$-body potentials, making them
comparable in efficiency to the corresponding $N$-body regular
potentials. In the following, for the sake of clarity we limit our
discussion to pair potentials. The extension to cluster potentials of
higher order is straightforward. 

For reference, let us first consider a conventional mixed pair
potential energy expression for a binary system,
\begin{equation}
E_{\text{pp}} = \sum_{i\in A}\sum_{j\in B} V(r_{ij}).
\end{equation}
Within this model the force on a particle $k$ of type $A$ is
calculated as follows
\begin{equation}
  \frac{\partial E_{\text{pp}}}{\partial \vec{r}_k^A} = 
  \sum_{j\in B} V'(r_{kj}) \frac{\vec{r}_{kj}}{r_{kj}}.
\end{equation}
Let us now consider a typical composition-dependent pair potential
model for the same binary system,
\begin{equation}
  E_{\text{cdpp}} = \sum_{i\in A}\sum_{j\in B} h(x^A_{ij})~V(r_{ij}),
\end{equation} 
where $x^A_{ij}$ is the two-centre concentration of the species $A$
about the $(i,j)$ pair. Now the force on particle $k$ of type $A$ can
be written
\begin{equation}
  \label{eq:force0}
  \frac{\partial E_{\text{cdpp}}}{\partial \vec{r}_k^A}  =
  \sum_{j\in B} V'(r_{kj}) h(x^A_{kj})  +
  \sum_{i\in A}\sum_{j\in B}V(r_{ij})h'(x_{ij}^A)\frac{1}{2}
  \left( \frac{\partial x^A_{i(j)}}{\partial \vec{r}_k^A} + 
  \frac{\partial x^A_{j(i)}}{\partial \vec{r}_k^A} \right), 
\end{equation}
for which after some algebra we obtain
\begin{equation}
  \frac{\partial x^A_{i(j)}}{\partial \vec{r}_k^A}
  = 
  \frac{\overline{\sigma}_i^B-\delta(\mathcal{S},t_j)\sigma(r_{ij})}{
    \left(\overline{\sigma}_i\right)^2-\sigma(r_{ij})}\sigma'(r_{ik})
  \frac{\vec{r}_{ki}}{r_{ki}}.
\end{equation}
All the quantities above have already been defined in
Eqs.~(\ref{eq:conc_ij}) and (\ref{eq:2cntr}). The second term in
\eq{eq:force0} contains contributions from two particles $i$ and $j$
to the forces on particle $k$. Hence composition-dependent pair
potentials lead to explicit three-body forces, which usually implies
significantly more expensive to calculations. However, we will now
show that in the case of expressions such as \eq{eq:force0} one can
regroup the terms in such a way as to speed up the calculation of
forces drastically. To this end, let us introduce a per-atom quantity
that for an atom of type $A$ reads
\begin{equation}
  M_{i\in A}^{\mathcal{S}} = \sum_{j\in B} V(r_{ij}) h'(x_{ij}^A)
  \frac{\overline{\sigma}_i^{\mathcal{S}}-\delta(B,t_j)\sigma(r_{ij})
  }{\left(\overline{\sigma}_i\right)^2-\sigma(r_{ij})},
\end{equation}
and for an atom of type $B$
\begin{equation}
  M_{i\in B}^{\mathcal{S}} = \sum_{j\in A} V(r_{ij}) h'(x_{ij}^A)
  \frac{\overline{\sigma}_i^{\mathcal{S}}-\delta(A,t_j)\sigma(r_{ij})
  }{\left(\overline{\sigma}_i\right)^2-\sigma(r_{ij})}.
\end{equation}
Substituting $M_i^{\mathcal{S}}$ into \eq{eq:force0} we obtain
\begin{equation}
  \label{eq:force1}
  \frac{\partial E_{\text{cdpp}}}{\partial \vec{r}_k^A}  =
  \sum_{j\in B} V'(r_{kj}) h(x^A_{kj})  +
  \frac{1}{2} \sum_i M_i^B \sigma'(r_{ki})\frac{\vec{r}_{ki}}{r_{ki}}.
\end{equation}
Similar derivation for the force on a particle $k$ of type $B$ leads
to the expression
\begin{equation}
  \label{eq:force2}
  \frac{\partial E_{\text{cdpp}}}{\partial \vec{r}_k^B}  =
  \sum_{j\in A} V'(r_{kj}) h(x^A_{kj})  +
  \frac{1}{2} \sum_i M_i^A \sigma'(r_{ki})\frac{\vec{r}_{ki}}{r_{ki}}.
\end{equation}
Each quantity in the above force expressions can be calculated
separately via pairwise summations. This allows for a very efficient
three-step algorithm for the calculation of forces:
({\it i}) compute and store the local partial densities
$\overline{\sigma}_i^{\mathcal{S}}$ for every atom,
({\it ii}) compute and store the quantities $M_i^{\mathcal{S}}$ for
every atom, and
({\it iii}) compute the forces according to the Eqs.~(\ref{eq:force1})
and (\ref{eq:force2}). This method leads to computational efficiency
comparable to standard EAM models.

\section{Linearised Models for efficient Monte-Carlo simulations}
\label{sect:monte_carlo}

Molecular dynamics simulations are limited when it comes to modelling
phenomena such as precipitation, surface and grain boundary
segregation, or ordering in alloys. Monte-Carlo (MC) methods, however,
are ideally suited for such applications. The most common techniques
are based on so-called swap trial moves, in which the chemical
identity of a random particle is changed. The resulting change in
potential energy, $\Delta E$, is used to decide whether the swap is
accepted or rejected.

The main task in an MC simulation is therefore to calculate the change
in potential energy induced by swapping the type of a single atom. For
short-range potentials this can be done very efficiently, since the
type exchange only affects the atoms in the neighbourhood of the type
swap. In the framework of the standard EAM model the situation is as
follows: Changing the species of one atom directly affects (1) its
embedding energy, (2) its pair-wise interactions with neighbouring
atoms, and (3) indirectly changes the electron density at neighbouring
atoms and therefore their embedding energies. All these quantities
need to be recalculated by visiting the atoms affected by the type
swap.

In the case of composition-dependent models the situation turns out
to be more laborious. To illustrate this let us again consider a
typical composition-dependent pair potential model for a binary
system:
\begin{equation}
E_{\text{cdpp}} = \sum_{i\in A}\sum_{j\in B} h(x^A_{ij})~V(r_{ij}),
\label{eq:cdpp1}
\end{equation} 
where $x^A_{ij}$ is the two-centre concentration of the species $A$
about the $(i,j)$ pair
\begin{equation}
  x^A_{ij}
  = \frac{1}{2}\left(x^A_{i(j)} + x^A_{j(i)} \right),
\end{equation}
where the $x^A_{i(j)}$ is the local concentration $A$ about the atom
$i$ excluding atom $j$. From \eq{eq:2cntr} we observed that  to a good
approximation $x_{i(j)}\approx x_i$. Therefore, for the qualitative
discussion below we replace $x_{i(j)}$ by $x_i$. In the energy
expression \eq{eq:cdpp1}, the site energy $E_i$ of an atom $i$ does
not only depend on the local concentration $x_i$, but also on the
concentrations $x_j$ of all its neighbours $j$. This has a dreadful
impact on the efficiency of the energy calculation. Changing the
chemical identity of some atom $i$ alters the local concentrations
$x_j$ of all its direct neighbours $j$, which in turn affects the mixed
interaction of all atoms $j$ with all of their respective neighbour
atoms $k$. All of these have to be re-evaluated to compute the total
change in energy induced by the single swap operation. The interaction
radius that has to be considered is therefore twice as large as the
cutoff radius of the underlying EAM potential, which increases the
computational costs by at least one order of magnitude.

This issue can be resolved quite effectively if we linearise the
interpolation function $h(x^A_{ij})$ as follows
\begin{equation}
  h(x^A_{ij}) = \frac{1}{2} \left(h(x^A_{i(j)}) + h(x^A_{j(i)}) \right).
\end{equation}
Within the new linearised formulation, although a single pair
interaction between two atoms $j$ and $k$ still depends on the
concentration at both sites, the site energy can be recast in a form
that is independent of the concentrations on the neighbouring sites. As
a result, the site energy of atom $k$ is no longer affected by
changing the type of an atom $i$ that is farther away than one cutoff
radius. Note that linearisation can be done for interpolation
functions of any $n$-centre concentrations. All composition-dependent
models independent of cluster size can therefore be linearised.
We have discussed the linearised model and its implementation for MD
and MC at length in a recent publication \cite{StuSadErh09}.

\section{A practical example}
\label{sect:FeCr}

To provide a practical illustration of the concepts developed in this
paper, we now revisit the composition-dependent EAM potential for
Fe--Cr \cite{CarCroCar05}, which has already been successfully applied
in a number of cases \cite{CarCarLop06b,ErhCarCar08}.

\subsection{Application of composition-dependent embedded atom method to Fe--Cr}

Iron alloys are materials with numerous technological applications. In
particular Fe--Cr alloys are at the basis of ferritic stainless
steels. It has been recently shown \cite{OlsAbrVit03} that the Fe--Cr
alloy in the ferromagnetic phase has an anomaly in the heat of
formation which shows  a change in sign going from negative to
positive at about 10\% Cr and leads to the coexistence of
intermetallic phase \cite{ErhSadCar08} and segregation in the same
alloy. This complexity results from a ``magnetic frustration'' of the
Cr atoms in the Fe matrix \cite{KlaDraFin06} which leads to an
effectively repulsive Cr-Cr interaction. Capturing this complexity
with an empirical potential model has been an active subject of
research in recent years.

To model this system, Caro and coworkers used the following ansatz
\begin{eqnarray}
  \label{eq:FeCr}
  E_{\text{Fe--Cr}} &=& \sum_{i\in {\text Fe}} U_{\text{Fe}} \left(\overline{\rho}_i^{\text{Fe}}+\overline{\rho}_i^{\text{Cr}}\right) + 
  \frac{1}{2} \sum_{i\in {\text {Fe}}}\sum_{j\in {\text{Fe}}}\phi_{\text{Fe}}\left(r_{ij}\right)\\
      &+&  \sum_{i\in {\text{Cr}}} U_{\text{Cr}}\left(\overline{\rho}_i^{\text{Cr}} + \overline{\rho}_i^{\text{Fe}}\right) + 
  \frac{1}{2} \sum_{i\in \text{Cr}}\sum_{j\in \text{Cr}}\phi_{\text{Cr}}\left(r_{ij}\right) ,\nonumber\\
  &+& \sum_{i\in \text{Fe}}\sum_{j\in \text{Cr}} h\left(\frac{x_i + x_j}{2}\right) V_{\text{mix}}(r_{ij}),\nonumber
\end{eqnarray}
where we used the same notation as in the earlier sections. The
partial electron densities $\overline{\rho}^{\mathcal{S}}_i$ follow
the same definition as in \eq{eq:rhobar}. Furthermore, the local
concentration variable $x_i$ in \eq{eq:FeCr} is defined as
\begin{equation}
  x_i = \frac{\overline{\rho}^{\text{Cr}}_i}{\rho^{\text{Cr}}_i+\rho^{\text{Fe}}_i}.
\end{equation}
The two densities $\rho^{\text{Fe}}(r_{ij})$ and
$\rho^{\text{Cr}}(r_{ij})$ are normalised such that at the equilibrium
lattice constant of each pure lattice, the respective partial electron
density is 1. In this way the two EAM models for the pure elements are
made compatible with each other.

Equation~(\ref{eq:FeCr}) looks quite similar to the
composition-dependent pair potential energy expressions discussed in
\sect{sect:pair_potentials}. There are, however, three essential
differences:
(i) There is only one mixed pair potential $V_{\text{mixed}}(r_{ij})$
as opposed to two in \sect{sect:pair_potentials} (one for each limit).
(ii) There is no boundary conditions on the interpolation function
$h(x)$ at $x=0$ and $x=1$.
(iii) The local concentration about the $(i,j)$ pair is just the
average of the one-centre concentrations about the two sites, and not
the two-centre concentration as defined in \eq{eq:2cntr}. Of course,
at no extra cost the more rigorous definition in \eq{eq:2cntr} is a
better choice for the measure of local concentration about a pair of
atoms.  On the other hand, \eq{eq:conc_ij} shows that the one-centre
concentration above is only a perturbation away from the more accurate
quantity.

The Fe--Cr CD-EAM model was the pioneering work that  has inspired the
current paper. Here, we have tried to give a more rigorous foundation
to the CD-EAM model. In fact, we can strictly argue that CD-EAM is a
simplified version of the current formalism. It works very well for
the Fe--Cr system since the two elements are similar in size and
chemical nature. It is therefore reasonable to make the approximation
that functional forms of the mixed pair potentials describing the two
dilute limits are the same.

Let us illustrate the last statement with the example of Lennard-Jones
(LJ) potentials. These potentials are determined by two parameters:
$\sigma$ and $\epsilon$; the first parameter specifies the position of
the minimum of the potential or in other words the particle size, and
the second parameter specifies the interaction strength. A mixture of
two types of LJ particles with no size mismatch (same $\sigma$) but
different cohesive energies can be described by the same potential
that is merely scaled differently for the two particles. Extending
this analogy to the case of the Fe--Cr system we can see why only one
mixed potential  can be enough. However, it is important to realise
now that when only one potential is used, the functions $h(x)$ provide
the interaction strength, which in the case of Fe--Cr is positive in
one dilute limit and negative in the other. Hence no boundary
conditions exist at the two concentrations $x=0$ and $x=1$.

In the original CD-EAM model, there was a further simplification. The
mixed potentials $V_{\text{mix}}(r_{ij})$ was never fitted. In fact it
was taken as the average of the effective EAM pairwise interactions of
the pure elements at their respective equilibrium volumes
\begin{equation}
V_{\text{mix}}(r_{ij})
= \frac{1}{2}\left( \phi_{\text{Fe}}(r_{ij}) + 2U_{\text{Fe}}(\overline{\rho}^{\text{Fe}}_0)\rho^{\text{Fe}}(r_{ij}) +  
\phi_{\text{Cr}}(r_{ij}) + 2U_{\text{Cr}}(\overline{\rho}^{\text{Cr}}_0)\rho^{\text{Cr}}(r_{ij}) \right),
\end{equation}
where $\overline{\rho}^{\mathcal{S}}_0$ is the electron density at the
equilibrium lattice constant for the species $\mathcal{S}$. Only the
function $h(x)$  was fitted to the heat of mixing of the solid
solution. The success of this model in spite of all the
simplifications is a telltale of the power of this methodology.

\subsection{Molecular dynamics and Monte Carlo performance}
\label{sect:MolecularDynamicsPerformance}

\begin{figure}
  \centering
\includegraphics[width=0.6\linewidth]{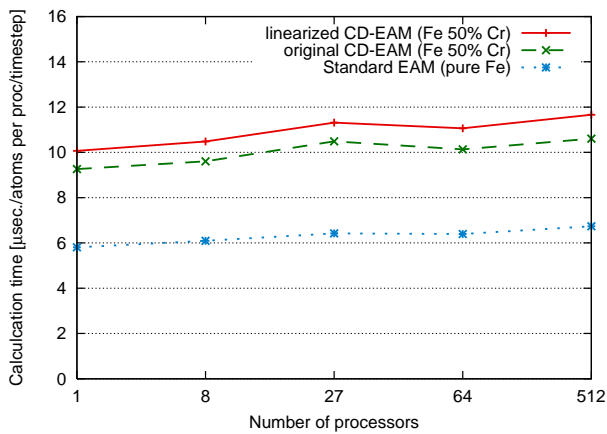}
  \caption{
    Comparison of the computation times for the CD-EAM models and the
    standard EAM model in a parallel molecular dynamics
    simulation. The benchmark simulation consists of a body-centred
    cubic crystal at 300\,K with 16{,}000 atoms per processor.
  }
  \label{fig:ScalingBenchmark}
\end{figure}

In \sect{sect:forces} we presented an algorithm for calculating forces
within the composition-dependent interatomic potential models which
brings their efficiency on par with the standard EAM scheme. This was
first discussed in a recent publication by the present authors
\cite{StuSadErh09}, where this algorithm was implemented for the
Fe--Cr CD-EAM model in the popular massively-parallel MD code LAMMPS
\cite{Pli95}.To benchmark its performance, we carried out MD
simulations of a body-centred cubic (BCC) crystal at 300\,K using
periodic boundary conditions. For the CD-EAM case we considered a
random alloy with 50\%\ Cr. For the standard EAM case, the sample
contained only Fe. Simulations were run on 1, 8, 27, 64, and 512
processors with 16,000 atoms per processor (weak scaling). The results
for the CD-EAM routines and the LAMMPS standard EAM routine are
displayed in \fig{fig:ScalingBenchmark}.
In this figure, the original CD-EAM model as well as its linearised
version are displayed. We see that the two versions are between 60\%
(linearised model) to 70\% (original model) slower than the standard
EAM. This is a small price to pay considering the fact that the CD-EAM
expression actually contains explicit three-body forces.

\begin{figure}
  \centering
\includegraphics[width=0.6\linewidth]{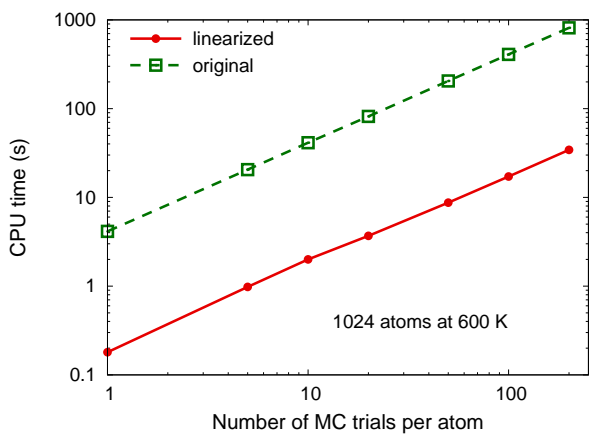}
  \caption{
    Comparison of the timing in a MC simulation of a Fe--Cr alloy at
    50\%\ composition. The simulation cell contained 1024 atoms.
  }
  \label{fig:mc_performance}
\end{figure}

In our recent publication \onlinecite{StuSadErh09} we also studied the Monte
Carlo performance of composition-dependent interatomic potentials focusing on
the comparison of the original and the linearised CD-EAM model. The
performance gain due to the linearised formulation is illustrated in
\fig{fig:mc_performance} which compares the timing of the linearised
and original CD-EAM models in a serial MC simulation for a random
Fe--Cr alloy at 50\%\ composition. We find that the linearized CD-EAM
model is twelve times faster than the original formulation. This is an
impressive performance gain, which clearly advocates for linearised
composition-dependent interatomic potentials.

\section{Conclusions}

\begin{figure}
  \centering
\includegraphics[scale=0.11]{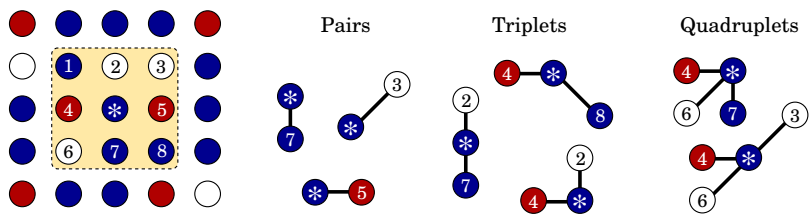}
  \caption{
    Several examples for clusters used to construct higher-order
    interaction terms which can be extracted from the configuration
    shown on the left.
  }
  \label{fig:schematic_clusters2}
\end{figure}

The present work has come about in response to a need for a practical
scheme for fitting interatomic potential models for multicomponent
alloys. At this point of time, when faced with the task of modelling
the chemistry of e.g. a ternary alloy, one is overwhelmed with the
complexity of the problem. In this paper, we have presented a
systematic methodology for the construction of alloy potentials,
starting from pre-existing potentials for the constituent
elements. The formalism represents a generalisation of the approach
employed by one of the authors for the Fe--Cr system
\cite{CarCroCar05}. We have shown that this formalism naturally
extends to treating multicomponent systems. The main idea of the
approach is to describe the energetics of dilute concentrations of
solute atoms in the pure host in terms of pair and higher-order
cluster interactions (see Figs.~\ref{fig:schematic_clusters1} and
\ref{fig:schematic_clusters2}). These interaction functions are then
used as a basis set for expanding the potential energy of the alloy
in the entire concentration range. To describe the energetics of the
concentrated alloys, the contributions of the basis functions are
weighted by interpolation functions expressed in terms of local
concentration variables. One of the innovations in this work is a
novel measure of local composition around individual atoms in the
system. This introduces an explicit dependence on the {\it chemical}
environment. In this sense the composition-dependent interatomic
potential scheme is reminiscent of the bond-order potential scheme
developed by Abell and Tersoff \cite{Abe85, Ter86, Ter88a} which
employs a measure of the bond-order to distinguish between different
{\it structural} motifs.

The main advantage of the framework presented here is that the basis
functions can be constructed sequentially and independent of the
interpolation functions, leading to a scheme that can be practically
implemented and systematically improved upon. The lower order terms
can be constructed with no knowledge of the higher order terms and
therefore need not be reparametrised when higher order cluster
potentials are constructed. The higher order terms in the expansion
become progressively smaller. In this way the model can be made step
by step, starting from the lowest order cluster
potentials. Furthermore addition of new terms in the series expansion
is not likely to introduce unphysical behaviour, a problem that plagues
most fitting schemes for interatomic potentials.

The practical determination of the basis functions and the
interpolation functions proceeds by fitting to first-principles
data. The expansion in cluster interactions may be reminiscent of
the celebrated ``cluster expansion'' technique \cite{SanDucGra84}
that has been used extensively during the past few decades to model
the thermodynamics of multicomponent alloys from first principles. But
it is important to note here that the methodology presented in this
paper has no relation to the cluster expansion technique. The latter
reduces the continuous phase space of e.g., a binary alloy onto the
discrete configuration space of the corresponding Ising model. There
is only one number associated with each cluster configuration, namely
the the free energy of that cluster. The so-called ``effective cluster
interactions'' (ECIs) are usually obtained via an optimisation process
from all the cluster free energies. A procedure of the sort proposed
in this paper is not possible, since there is not direct link between
any single cluster free energy and an ECI. In contrast, when fitting
e.g. a $V_{AB}(r_{ij})$ interaction potential, a solute inclusion not
only changes the total energy of the system, it causes forces in the
system and modifies the force constants of the host, all of which can
be used to construct a continuous pair potential.

Composition-dependent interatomic potentials are constructed by
incorporating pair, triplet and higher-order cluster interactions
that describe the energetics of clusters embedded in a pure host with
a specific underlying lattice. One may now wonder, with this approach,
could a potential be expected to handle systems which 
change lattice type as a function of concentration? For instance the
Ni-Al phase diagram contains phases with BCC-based crystal structures,
while the pure metals are face-centred cubic (FCC). Following the
approach described above, the basis functions are parametrised in
terms of solute cluster energies in the constituent FCC
structures. How can one then expect to provide a reasonable model for
the BCC-based NiAl phase? The answer lies in the interpolation
functions.They are fitted to the energetics of the ordered and
disordered compounds along the concentration range with arbitrary
crystal structures.

\section*{Acknowledgements}
Lawrence Livermore National Laboratory is operated by Lawrence
Livermore National Security, LLC, for the U.S. DOE-NNSA under Contract
DE-AC52-07NA27344. Partial financial support from the LDRD office and
the Fusion Materials Program as well as computer time allocations from
NERSC at Lawrence Berkeley National Laboratory are gratefully
acknowledged.

\bibliographystyle{unsrt}

\end{document}